\documentclass[aps,prl,twocolumn,showpacs]{revtex4}
\usepackage{graphicx}
\usepackage{color}
\usepackage{bm}

\begin{document}

\title{Switching superconductivity  in S/F bilayers by multiple-domain structures
}
\author{T. Champel and M. Eschrig}
\affiliation{
Institut f\"{u}r Theoretische Festk\"{o}rperphysik,
  Universit\"{a}t Karlsruhe,
 76128 Karlsruhe, Germany }
\date{\today}

\begin{abstract}
We consider the effect of a multiple magnetic domain structure in
a superconductor/ferromagnet bilayer, modeled by a ferromagnetic 
layer with a rotating magnetic moment. The domain walls in this model
are of equal size as the domains, and are of N\'{e}el type.
We study the superconducting critical temperature as a function of
the rotation wavelength of the magnetic moment.
The critical temperature of the bilayer is found to 
be always enhanced by the domain structure, 
and exhibits an interesting reentrant 
behavior. We suggest that this effect can be used for a
new device where superconductivity may be controlled by 
the domain structure of the magnetic layer.
\end{abstract}


\maketitle
The study of the proximity effect between a superconductor (S) and a 
ferromagnet (F) is currently a very active field due to its relevance 
for nanoelectronic applications and due to the perspectives 
to discover new interesting physical phenomena. 
It is well-known  that the exchange field ${\bf J}$ of the 
ferromagnet tends to break the  Cooper pairs formed by electrons with 
opposite spins  by acting on the electronic spins via the exchange
interaction. 
When the exchange field exceeds the typical superconducting low-energy scales,
$J > 2\pi T_{c0}$ ($T_{c0}$ is the superconducting  critical temperature in absence of the proximity effect), a new, shorter length scale competes with the 
superconducting coherence length scale.
For this case,
within the quasi-classical theory of superconductivity for 
diffusive structures (Usadel equations \cite{Usa1970}), 
the superconducting pair correlations have been 
found to penetrate (and also oscillate) in the F part over a short length 
$\xi_{J}=\sqrt{D_{f}/J}$ instead of the superconducting coherence length 
$\xi_{f}=\sqrt{D_{f}/2 \pi T_{c0}}$
($D_{f}$ is the diffusion constant in F) \cite{Rad1991}.
Moreover, in S/F bilayers 
the critical temperature $T_{c}$ is suppressed by the proximity
of the F layer and exhibits 
a non-monotonic dependence on the thickness of the F layer, $d_{f}$
(see \cite{Fom2002} and references therein). 
In F/S/F sandwiches, it has been found that the pair-breaking effect depends 
on the relative orientation of the F moments. This property led to the proposal of a superconducting switch \cite{Tag1999,Buz1999} operated by reversing the moment in a F layer.

In this Letter, we introduce a new aspect of this feature. We investigate
the sensitivity of the Cooper pairs on the directional changes of the 
exchange field by studying  a S/F bilayer with a moment rotating with a 
constant velocity in the F layer \cite{Ber2000} (see Fig. 1). 
\begin{figure}[t]
\begin{center}
\begin{picture}(0,0)%
\includegraphics{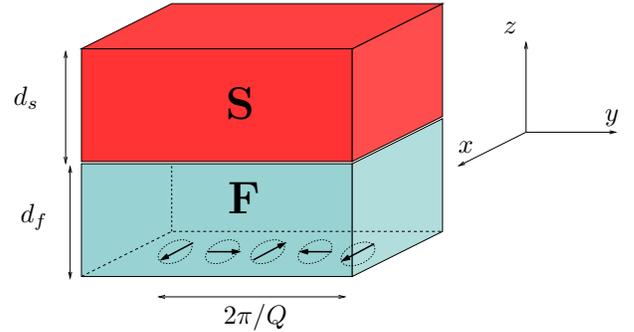}%
\end{picture}%
\setlength{\unitlength}{1243sp}%
\begingroup\makeatletter\ifx\SetFigFont\undefined
\def\x#1#2#3#4#5#6#7\relax{\def\x{#1#2#3#4#5#6}}%
\expandafter\x\fmtname xxxxxx\relax \def\y{splain}%
\ifx\x\y   
\gdef\SetFigFont#1#2#3{%
  \ifnum #1<17\tiny\else \ifnum #1<20\small\else
  \ifnum #1<24\normalsize\else \ifnum #1<29\large\else
  \ifnum #1<34\Large\else \ifnum #1<41\LARGE\else
     \huge\fi\fi\fi\fi\fi\fi
  \csname #3\endcsname}%
\else
\gdef\SetFigFont#1#2#3{\begingroup
  \count@#1\relax \ifnum 25<\count@\count@25\fi
  \def\x{\endgroup\@setsize\SetFigFont{#2pt}}%
  \expandafter\x
    \csname \romannumeral\the\count@ pt\expandafter\endcsname
    \csname @\romannumeral\the\count@ pt\endcsname
  \csname #3\endcsname}%
\fi
\fi\endgroup
\begin{picture}(12027,6447)(451,-6946)
\put(586,-4876){\makebox(0,0)[lb]{\smash{\SetFigFont{10}{12.0}{rm}{\color[rgb]{0,0,0}$d_{f}$}%
}}}
\put(451,-2536){\makebox(0,0)[lb]{\smash{\SetFigFont{10}{12.0}{rm}{\color[rgb]{0,0,0}$d_{s}$}%
}}}
\put(4726,-4696){\makebox(0,0)[lb]{\smash{\SetFigFont{17}{20.4}{rm}{\color[rgb]{0,0,0}\textbf{F}}%
}}}
\put(4681,-2806){\makebox(0,0)[lb]{\smash{\SetFigFont{17}{20.4}{rm}{\color[rgb]{0,0,0}\textbf{S}}%
}}}
\put(4636,-6901){\makebox(0,0)[lb]{\smash{\SetFigFont{10}{12.0}{rm}{\color[rgb]{0,0,0}$2 \pi /Q$}%
}}}
\put(10216,-1096){\makebox(0,0)[lb]{\smash{\SetFigFont{10}{12.0}{rm}{\color[rgb]{0,0,0}$z$}%
}}}
\put(9316,-3481){\makebox(0,0)[lb]{\smash{\SetFigFont{10}{12.0}{rm}{\color[rgb]{0,0,0}$x$}%
}}}
\put(12241,-2851){\makebox(0,0)[lb]{\smash{\SetFigFont{10}{12.0}{rm}{\color[rgb]{0,0,0}$y$}%
}}}
\end{picture}
\vspace{-0.2cm}
\caption{Model: the moment ${\bf J}$ rotates in the F layer.}
\vspace{-1cm}
\end{center}
\end{figure}
This model simulates a multi-domain structure with domain walls of the
N\'{e}el type, where the domain wall width is comparable in size to 
the size of the domains.
We find that the critical temperature $T_{c}$ is enhanced compared to 
that of the S/F bilayer with homogeneous magnetization
and exhibits an interesting reentrant behavior. 
We suggest to use this effect for new devices where superconductivity 
may be switched on and off by controlling the magnetic domain structure.

Recently, the theoretical understanding of the interplay between the 
superconductivity and inhomogeneous magnetism has been improved by
realizing that a local inhomogeneous exchange field induces superconducting 
triplet correlations, which penetrate over a long scale, $\xi_{f}$,  
in a diffusive ferromagnet \cite{Ber2001,Kad2001}. 
As a result, these correlations lead to a significant increase of the 
conductance of the ferromagnet below $T_{c}$ \cite{Ber2001}.
Volkov {\em et al.} \cite{Vol2003,Ber2003} have demonstrated the 
existence of these long-range triplet components (TC) also in the 
multilayered F/S/F structures  when the moments in each F layers are 
non-collinear.
As already emphasized in Ref. \cite{Ber2001,Vol2003,Ber2003}, these TC have the property to be odd in frequency, a necessary condition for their existence 
in the diffusive limit.
The study of $T_{c}$ as a function of the angle between the moments in 
the F/S/F trilayer has been reported very recently \cite{Fom2003}.
In order to clarify the situation and to compare with our model,
we first derive generally the Usadel equations in the paired-spin basis
and show that the coexistence  of singlet components (SC) with 
odd-triplet components is in fact the hallmark of any diffusive S/F structure \cite{Rem}. 
Short-range TC are necessarily present. Contrarily, the presence of 
TC with long-range scales requires specific conditions.
For example, the long-range TC are shown to be absent within our model.
The sensitivity of $T_{c}$ on the magnetic inhomogeneity is 
in our model mainly 
provided by the short-range TC.

We study the S/F bilayer within the framework of the Nambu-Gor'kov formalism 
and consider the diffusive limit which is relevant for short mean free paths.
The Usadel Green's function $\hat{g}({\bf R}, \omega_{n})$ depends on the 
spatial coordinate ${\bf R}$ and on the Matsubara frequency $\omega_{n}=\pi T 
(2 n+1)$ (where $T$ is the temperature), and results from the momentum average 
of the quasi-classical Green's function. 
It is represented by a $4 \times 4$ Nambu-Gor'kov matrix in combined 
particle-hole and spin spaces.
In particle-hole space it is written as
\begin{equation}
\label{Eq1}
\hat{g}=\left( \begin{array}{cc}
g & f\\
f^{\dag} & g^{\dag}
\end{array}
\right)
\end{equation}
where the 2x2 spin-matrices $g$ and $f$ represent the normal and anomalous 
Green's functions.
$\hat{g}$ obeys  a normalization condition and a nonlinear transport equation,
coupling its components $f$ and $g$ \cite{Usa1970}.
Near $T_{c}$, $f$ is small and $g$ deviates slightly from its value ($-i \pi \mathrm{sgn} \, \omega_{n}$) 
in the normal state, so that the Usadel transport Eq. can be 
linearized with the help of the normalization condition, and yields an
equation for the 2x2 spin-matrix $f$
\begin{eqnarray}
\label{Eq2}
D \bm{\nabla}^{2}f 
-i \,
\mathrm{sgn} \,(\omega_{n}) \, 
\left\{{\bf J}\left({\bf R}\right) \cdot \bm{\sigma} f+f{\bf J}\left({\bf R}\right) \cdot \bm{\sigma}^{\ast} \right\} \nonumber \\
-2|\omega_{n}|f 
+2 \pi \Delta =0
\end{eqnarray}
with $\bm{\sigma}=(\sigma_{x},\sigma_{y},\sigma_{z})$ a vector whose 
components are the Pauli matrices. 
In the F layer the superconducting order parameter $\Delta$ vanishes,
and in the S layer the exchange field ${\bf J}$ is zero. 
The diffusion constants are $D=D_{s}$ in S and $D=D_{f}$ in F.
We consider a BCS singlet superconductor for which the order parameter in 
spin-space is
$\Delta= \Delta_{s} i \sigma_{y}$.
Quite generally, $f$ is expected to have a 
singlet and a triplet part \cite{Min1999} 
\begin{equation}
\label{Eq4}
f= f_{s} i \sigma_{y} + i \left({\bf f}_{t}\cdot \bm{\sigma}\right) \sigma_{y} 
\end{equation}
where ${\bf f}_{t}=(f_{tx},f_{ty},f_{tz})$ is the triplet vector 
(the basis formed by the vectors $s=(\left| \uparrow \downarrow \right\rangle- \left| \downarrow \uparrow \right\rangle)/2$, 
$t_{x}=(\left| \downarrow \downarrow \right\rangle- \left| \uparrow \uparrow \right\rangle)/2$,
$t_{y}=(\left| \uparrow \uparrow \right\rangle+\left| \downarrow \downarrow \right\rangle)/2i$, and $t_{z}=(\left| \uparrow \downarrow \right\rangle+\left| \downarrow \uparrow \right\rangle)/2$ is the natural basis for studying 
pairing states).
The different components of 
$f$ obey the system of coupled equations
\begin{eqnarray}
\left(D \bm{\nabla}^{2}-2|\omega_{n}| \right)&f_{s}=& -2\pi \Delta_{s}+2i\, \mathrm{sgn}(\omega_{n}) \, {\bf J} \cdot {\bf f}_{t} \label{Eq5}
\\
\left(D \bm{\nabla}^{2}-2|\omega_{n}| \right)&{\bf f}_{t}=& 2i \, \mathrm{sgn}(\omega_{n}) \,
{\bf J}f_{s}. \label{Eq6}
\end{eqnarray}

The couplings (right-hand side) in Eq. (\ref{Eq5})-(\ref{Eq6}) show 
obviously that the SC $f_{s}$ coexists always with at least one 
TC as early as ${\bf J} \neq {\bf 0}$. 
We see also generally that the SC is even in frequency 
[$f_{s}(-\omega_{n})=f_{s}(\omega_{n})$] as it is the case for the 
BCS superconductors, while the TC are odd [
${\bf f}_{t}(-\omega_{n})=-{\bf f}_{t}(\omega_{n})$]. 
This is a consequence of the Pauli principle, which for the 
diffusive limit  imposes the relation $f_{\alpha \beta}({\bf R},\omega_{n})=-f_{\beta \alpha}({\bf R},-\omega_{n})$, where $\alpha$ and $\beta$ are the spin indices.
Henceforth, we only consider positive Matsubara frequencies.
The Eq. (\ref{Eq5})-(\ref{Eq6}) are supplemented by the self-consistent equation relating $\Delta_{s}$ to $f_{s}$ 
\begin{equation}
\label{Eq7}
\Delta_{s}\ln \frac{T_{c0}}{T}=2\pi T \sum_{n \geq 0} \left(
\frac{\Delta_{s}}{\omega_{n}}-
\frac{f_{s}(\omega_{n})}{\pi}
\right), \label{self}
\end{equation}
and also by the boundary conditions. The general boundary conditions at the
S/F interface for the diffusive case have been formulated by 
Nazarov \cite{Naz99} and reduce near $T_{c}$ to
\begin{equation}
\label{Eq8}
\xi_{s} \left. \partial_{z} f\right)_{S}=
\gamma \xi_{f} \left. \partial_{z} f\right)_{F}, \hspace{0.5cm} \gamma=\rho_{s} \xi_{s}/\rho_{f}\xi_{f}
\end{equation}
where 
$\rho_{s}$ and $\rho_{f}$ are respectively the normal-state resistivities of the S and F metals
[Eq. (\ref{Eq8}) follows from the continuity of the current at the interface], and
\begin{equation}
\label{Eq9}
\xi_{f} \gamma_{b} \left. \partial_{z} f \right)_{F}= \left. f \right)_{S}-\left. f \right)_{F}, \hspace{0.5cm} \gamma_{b}=R_{b} {\cal A}/\rho_{f} \xi_{f}
\end{equation}
with $R_{b}$ the resistance of the S/F boundary, and ${\cal A}$ its area. Here $z$ denotes the distance to the S/F interface. 
Note that near $T_c$ Nazarov's boundary conditions are formally equivalent 
to the ones by Kuprianov and Lukichev~\cite{Kup1988}.
At the outer surfaces of the F or S layers, we require that the current through the boundary has to vanish.
It is important to note that 
the conditions (\ref{Eq8})-(\ref{Eq9})
do not couple the different components  of $f$. 

If ${\bf J}$ is constant in direction, we obtain straightforwardly that the triplet vector ${\bf f}_{t} \parallel {\bf J}$.
This configuration  corresponds to a triplet state with a zero spin 
projection on the quantization axis defined by $\hat{{\bf J}}$. 
The SC and the TC with zero spin projection are 
energetically equivalent with respect to the exchange interaction,
and thus necessarily appear together in the ferromagnet.

In F/S/F trilayers  with an arbitrary angle between the moments in each 
F layer \cite{Vol2003,Ber2003,Fom2003}, ${\bf J}$  has not a fixed direction in the structure.
If the moments are homogeneous in each F layer, the system (\ref{Eq5})-(\ref{Eq6}) 
can be solved easily.
In the different F layers, the spatial dependence of the components of 
$f$ can be written as $f_{i}(z)=f_{i}\exp(k_{f} z)$, leading to
the algebraic equations 
\begin{equation}
\label{Eq10}
\left(\begin{array}{cc}
D_{f} k_{f}^{2}-2\omega_{n} & - 2i \, {\bf J} \cdot \\
-2 i \, {\bf J} & D_{f} k_{f}^{2}-2\omega_{n}
\end{array}
\right)
\left(\begin{array}{c}
f_{s}\\
{\bf f}_{t}
\end{array}
\right)
=0.
\end{equation}
The eigenvalue $k_{f}^{2}$ is determined from the condition of zero determinant, which yields
\begin{equation}
\label{Eq11}
k^{2}_{f\pm}=\Omega_{n}\xi_{f}^{-2} \pm 2 i \xi_{J}^{-2} \hspace{0.3cm} \mathrm{or} \hspace{0.3cm} k_{f0}^{2}=\Omega_{n} \xi_{f}^{-2} \label{eigen}
\end{equation}
with $\Omega_{n}=\omega_{n}/\pi T_{c0}$.
The corresponding eigenvectors have the form
\begin{equation}
\label{Eq12}
\left(\begin{array}{c}
f_{s \pm}\\
{\bf f}_{t\pm}
\end{array}
\right)=f_{s \pm}\left(\begin{array}{c}
1\\
\pm \hat{{\bf J}}
\end{array}
\right) 
\hspace{0.3cm} \mathrm {and} \hspace{0.3cm}
\left(\begin{array}{c}
f_{s 0}\\
{\bf f}_{t0}
\end{array}
\right)=\left(\begin{array}{c}
0\\
{\bf f}_{t0}
\end{array}
\right)
\end{equation}
with the condition ${\bf f}_{t0} \cdot {\bf J}=0$.
Thus,
there can exist {\em a priori} two very different types of 
TC \cite{Ber2001,Kad2001}.
The SC $f_{s}$ and the TC $\parallel{\bf J}$ characterized by $k_{f \pm}^{2}$ have the usual short-range 
decay length $\xi_{J}$ in the F layer for strong $J$. 
On the contrary, the TC ${\bf f}_{t} \perp {\bf J}$ (characterized by $k_{f0}^{2}$) penetrate further in the 
ferromagnet. Since they correspond (locally) to triplet Cooper pairs with a 
non-zero spin projection (equal-spin pairing), they are not locally
limited by the paramagnetic interaction with ${\bf J}$.

Now, we consider the S/F bilayer shown in Fig. 1 where ${\bf J}$ 
rotates  with a constant velocity $Q$ in the F layer, i.e. 
${\bf J}(y)=J ( \cos Q y, \sin Qy,0)$ (henceforth $z$ designates no longer the quantization axis). 
It is straightforward to see that here $f_{tz}=0$.  
The $y$ dependence of ${\bf J}$ is eliminated in the right-hand side 
of the Eq. (\ref{Eq5})-(\ref{Eq6}) by considering the new components 
$f_{+}=(-f_{tx}+if_{ty})e^{i Qy}$
and 
$f_{-}=(f_{tx}+if_{ty})e^{-i Qy}$. The system of Eq. to solve is 
\begin{eqnarray}
\label{Eq13}
\left(D\bm{\nabla}^{2} -2\omega_{n}\right)
f_{s}=- 2 \pi \Delta_{s}+
 i \,J\left(f_{-}-f_{+}\right) 
\\
\label{Eq14}
\left(D\bm{\nabla}^{2} \mp 2 i D Q \partial_{y}
-D Q^{2} - 2\omega_{n}\right)f_{\pm}= \mp \, 2 i \, J f_{s} 
.
\end{eqnarray}
Since the structure is periodical in the $y$ direction, 
the three components $f_{l}$ ($l=s,\pm$) and $\Delta_{s}$ are expanded into Fourier series. 
We note that the different Fourier components (labelled by $p$) are neither mixed by the full system of linearized equations nor by the boundary conditions. Thus, each harmonic  taken separately (and determining a particular $y$ dependence) represents a possible solution. The harmonic $p$ realized physically is the one which gives the highest $T_{c}$. 
As a result of our calculations \cite{renvoi}, 
the harmonic $p=0$  yields the highest $T_{c}$. 
The physical reason for this is that the other harmonic solutions 
correspond to an inhomogeneous bulk superconducting state, 
which decreases unavoidably $T_{c}$. We present here the equations for the case $p=0$.

In the F layer, 
using  the boundary condition at the outer surface ($z=-d_{f}$), the components $f_{l}$ are thus sought under the form  $
f_{l}(z)= f_{l} \cosh\left[k_{f} (z+d_{f}) \right]$, which is substituted in the set of Eq. (\ref{Eq13})-(\ref{Eq14}). This leads to the following  linear system
\begin{equation}
\label{Eq15}
\left(\begin{array}{ccc}
\lambda & -i \xi_{J}^{-2} & i \xi_{J}^{-2}  \\
-2i \xi_{J}^{-2} & \lambda-Q^{2}& 0 \\
2 i \xi_{J}^{-2} & 0& \lambda-Q^{2}
\end{array}
\right)
\left(\begin{array}{c}
f_{s}\\
f_{-}\\
f_{+}
\end{array}
\right)
=0
\end{equation}
with the eigenvalue $\lambda=k^{2}_{f}-\Omega_{n}\xi_{f}^{-2}$. This system yields the three eigenvalues (and consequently three $k_{f}$)
\begin{equation}
\label{Eq16}
\lambda_{\varepsilon}=\left(Q^{2}+\varepsilon \sqrt{Q^{4}-16 \xi_{J}^{-4}} \right)/2, \hspace{0.7cm} \lambda_{0}=Q^{2}
\end{equation}
and the associated eigenvectors (here $\varepsilon= \pm 1$)
\begin{equation}
\label{Eq17}
\left( \begin{array}{c}
f_{s, \varepsilon} \\
f_{-, \varepsilon}\\
f_{+, \varepsilon}
\end{array}
\right)= 
\left( \begin{array}{c}
\lambda_{-\varepsilon} \\
-2 i \xi_{J}^{-2} \\
2 i \xi_{J}^{-2}
\end{array}
\right)
, \hspace{0.5cm} 
\left( \begin{array}{c}
f_{s, 0} \\
f_{-, 0}\\
f_{+, 0}
\end{array}
\right)=
\left( \begin{array}{c}
0\\
1 \\
1
\end{array}
\right).
\end{equation}
The eigenvalues $\lambda_{\varepsilon}$ give short-range decay lengths for the TC in strong ferromagnets, while the eigenvalue $\lambda_{0}$ determines a long-range penetration in the ferromagnet (at least at small $Q$).
The solution of Eq. (\ref{Eq13})-(\ref{Eq14}) independent of $y$ is thus written as
\begin{eqnarray}
\label{Eq18}
f_{l}(z)= \sum_{j=0, \varepsilon} a_{j} f_{l,j} \cosh\left[k_{fj} (z+d_{f}) \right]
\end{eqnarray}
where the three coefficients $a_{j}$  have to be determined with the help of the boundary conditions at the S/F interface.

In the S layer, the solutions for the TC $f_{\pm}$ satisfying the boundary condition at the outer surface (at $z=d_{s}$) are straightforwardly derived 
$ f_{\pm }(y,z)= c_{\pm}
\cosh\left[k_{s}(z-d_{s})\right] $
where  $
k_{s}=\sqrt{\Omega_{n} \xi_{s}^{-2}+Q^{2}}$,  $\xi_{s}=\sqrt{D_{s}/2 \pi T_{c0}}$, and $c_{\pm}$ are coefficients.
As a result of the form of the eigenvector associated with the eigenvalue $\lambda_{0}$, 
the boundary conditions for the TC at the S/F interface require
$a_{0}=0$. This absence of the long-range TC in the present model is characterized by the fact that the triplet vector ${\bf f}_{t}(y,z)$ 
is locally parallel to ${\bf J}(y)$.

The remaining equation for the SC $f_{s}(z)$  can generally not be  solved analytically in the S layer since it is coupled with the self-consistent gap $\Delta_s(z)$ by Eq. (\ref{self}).
We use a technical high-energy cutoff of $\omega_n=21\pi T_{c0}$ for the gap equation.
We iterate Eqs. (\ref{Eq13}) and (\ref{self}) in S numerically, 
together with the
boundary conditions (\ref{Eq8})-(\ref{Eq9}), which determine the coefficients
$a_{j}$ and $c_{\pm}$. 

\begin{figure}[t]
\begin{center}
\includegraphics[width=7.5cm]{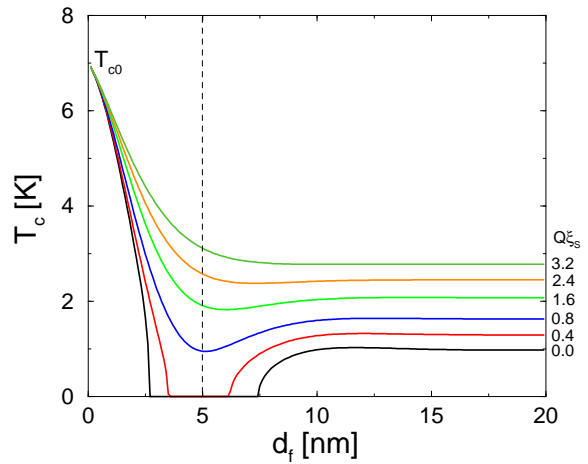}
\vspace{-0.4cm}
\caption{$T_{c}$ versus $d_{f}$ for various $Q \xi_{s}$. Here $d_{s}=7$ nm.
For $d_{f}\approx 5$ nm, corresponding to the dashed line, 
a switching effect can be observed between $Q=0$ (single domain F) and
$Q \xi_{s} > 0.56$.
}
\vspace{-0.6cm}
\end{center}
\end{figure}
In Fig. 2 we show the dependence of $T_{c}$ on $d_{f}$ for different values of the dimensionless parameter $Q \xi_{s}$ (for definiteness
we took the same parameters as in Fig. 2 of Ref. \cite{Fom2002}: $T_{c0}= 7$ K, $\gamma=0.15$, $\gamma_{b}=0.3$, $J=130$ K, $\xi_{s}=8.9$ nm, and $\xi_{f}=7.6$ nm).
$T_{c}$ clearly increases with $Q$. 
Moreover, we see that the non-monotonic behavior of 
$T_{c}(d_{f})$, responsible for the absence of superconductivity 
in a finite intermediate range of $d_{f}$ (around $d_f =$ 5 nm in
Fig. 2), tends to be suppressed in favor of a monotonic behavior. 
This tendency is accompanied by a
reentrance of superconductivity as a function of magnetic inhomogeneity
in the  $d_{f}$ range where $T_{c
}$ is zero. 
Fixing $d_f$ in this range we predict that when $Q$ exceeds some critical 
value $Q_{crit} $ ($\approx 0.56/\xi_s$ for $d_f = 5$ nm),
superconductivity is recovered.
By controlling the degree of magnetic inhomogeneity it is possible
to {\it switch the bilayer between the normal state and the 
superconducting state}.

The critical behavior is illustrated for $d_f=5$ nm in Fig. 3,
where $T_{c}$  is plotted as a function of $Q \xi_{s}$ for different $d_{s}$. 
It is clear that $T_c$ is enhanced for any $Q$ compared to the
homogeneous case $Q=0$.
\begin{figure}[t]
\begin{center}
\includegraphics[height=6cm]{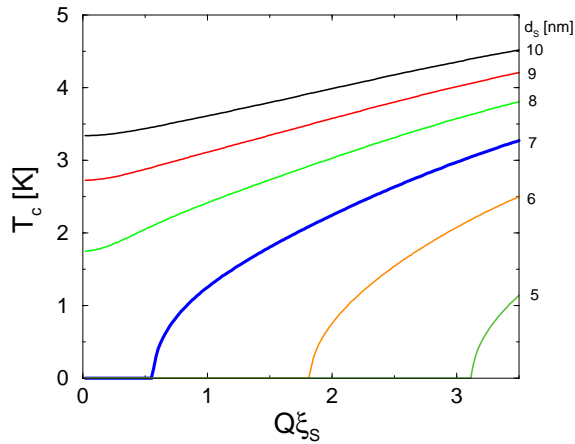}
\vspace{-0.4cm}
\caption{$T_{c}$ versus $Q \xi_{s}$ for $d_{f}=5$ nm and various $d_{s}$.
The thick curve corresponds to a cut along the dashed line in Fig. 2.
The onsets for $d_s < $ 7.44 nm behave like $\sqrt{Q-Q_{crit}}$.}
\vspace{-0.5cm}
\end{center}
\end{figure}
The switching behavior can be observed 
for superconducting layers with $d_s < d_{s,crit}$, where for the used 
parameters $d_{s,crit} \approx 7.44 $ nm. 
For $d>d_{s,crit}$ we observe for small $Q$ a behavior $T_c\sim Q^2$.
For $d<d_{s,crit}$ there is an onset of superconductivity at $Q_{crit}$, and
near the onset the functional behavior of $T_c$ as function of $Q$
is $T_c \sim \sqrt{Q-Q_{crit}}$.
By choosing the thickness of the superconducting layer appropriately,
it is possible to optimize the superconducting switch for the
corresponding range of $Q$ in which it can be controlled.

Finally, we discuss the implications of our results for the currently
debated general issue of the influence on the superconductivity 
of domain walls present in the F layer.
The evolution of ${\bf J}$ in the domain walls can typically be 
described by two kinds of models. In bulk material, ${\bf J}$ rotates 
about an axis perpendicular to the wall, 
i.e. it remains in the plane of the domain wall (Bloch wall).
On the contrary, in thin F layers ${\bf J}$ is expected to 
evolve in the surface plane (N\'{e}el wall). 
Our model can be seen as a model for a F layer containing many 
N\'{e}el walls whose widths are comparable to the widths of the domains.
Recently, a dependence of $T_{c}$ on the (controlled) 
domain state of the ferromagnet has been observed \cite{Rus} in S/F bilayers. 
In the presence of many domain walls, $T_{c}$ has been found to be slightly 
enhanced compared to the $T_{c}$ obtained in the absence of domain walls. 
The authors of the Ref. \cite{Rus} argue that the domain walls in their 
experiment are likely of the N\'{e}el type.
Indeed, the presence of Bloch walls  in a F layer induces 
unavoidably a non-zero component for ${\bf J}$ in the direction 
perpendicular to the layer. As a result, the exchange field acting on the 
electron spins produces in addition an electromagnetic field acting via 
the electronic charges on the orbital motion of electrons and thus 
suppresses the superconductivity. This rising orbital pair-breaking 
effect competes with the weakened paramagnetic pair-breaking effect 
in the domain walls and is expected to reduce or even prevent any 
dependence of $T_{c}$ on the domain state.
Our results are thus consistent with a $T_{c}$ enhancement due to 
N\'{e}el domain walls. A future step is to take into account 
more realistically the alternation of domain regions with domain walls. 
The difficulty is that the problem is intrinsically two-dimensional, and  
the solutions with separated $y$ and $z$ dependences are no more 
physically acceptable for boundary reasons. This prevents any simple 
analytical treatment even in the F layer as done here. 

The possibility to reversibly create and remove N\'{e}el walls in a controlled way as needed for the realization of the suggested device has been shown very recently \cite{Bun2005}. The method used in Ref. \cite{Bun2005} is based on the exchange bias effect between a ferromagnetic Co thin film and an antiferromagnetic insulating CoO film.

In conclusion, 
we have presented a model for a multiple magnetic domain structure in 
an S/F bilayer and have shown that
the superconducting critical temperature is enhanced by the magnetic
inhomogeneity and even exhibits 
reentrant superconducting behavior appealing for future applications.
We suggest to use this effect for a superconducting switch operated by
controlling the degree of inhomogeneity in the ferromagnetic layer.

This work was supported by the Deutsche Forschungsgemeinschaft within
the Center for Functional Nanostructures.


\begin{thebibliography}{99}
\bibitem{renvoi}
The technical details will be reported elsewhere.
\end{thebibliography}
\end{document}